\documentclass[10pt, conference, letterpaper]{IEEEtran}
\IEEEoverridecommandlockouts
\usepackage{cite}
\usepackage{amsmath,amssymb,amsfonts}
\usepackage{graphicx}
\usepackage{textcomp}
\usepackage{xcolor}
\usepackage [english]{babel}
\usepackage [autostyle, english = american]{csquotes}
\MakeOuterQuote{"}
\usepackage{enumitem}
\usepackage{graphicx}
\usepackage{float}
\usepackage{dsfont}
\usepackage{graphics} 
\usepackage{tabularx}
\usepackage{placeins}
\usepackage{multicol}
\usepackage{multirow}
\usepackage{tabularx}
\usepackage{pdflscape}
\usepackage{bbding}
\usepackage{subfigure}
\usepackage{afterpage}
\usepackage{rotating}
\usepackage{lscape}
\usepackage{booktabs}
\usepackage[noindentafter]{titlesec}
\usepackage{hyperref}
\graphicspath{ {images/} }
\usepackage{algorithm}
\usepackage{xcolor}
\usepackage{mathtools}
\DeclarePairedDelimiter\norm{\lVert}{\rVert}
\usepackage[noend]{algpseudocode}

\usepackage{mathtools,bm}

\DeclareMathOperator*{\argmin}{arg\,min}

\def\BibTeX{{\rm B\kern-.05em{\sc i\kern-.025em b}\kern-.08em
    T\kern-.1667em\lower.7ex\hbox{E}\kern-.125emX}}
    
\makeatletter
\patchcmd{\@maketitle}
  {\addvspace{0.5\baselineskip}\egroup}
  {\addvspace{-1\baselineskip}\egroup}
  {}
  {}
\makeatother

\begin{document}
\bstctlcite{IEEEexample:BSTcontrol}
\urlstyle{tt}
\title{MetaNet: Automated Dynamic Selection of Scheduling Policies in Cloud Environments}

\author{
\IEEEauthorblockN{Shreshth Tuli\IEEEauthorrefmark{1}, Giuliano Casale\IEEEauthorrefmark{1}, Nicholas R. Jennings\IEEEauthorrefmark{1}\IEEEauthorrefmark{2}}
\IEEEauthorblockA{\IEEEauthorrefmark{1}{Imperial College London}}
\IEEEauthorblockA{\IEEEauthorrefmark{2}{Loughborough University}}
\IEEEauthorblockA{\{s.tuli20, g.casale\}@imperial.ac.uk, n.r.jennings@lboro.ac.uk}%
}

\maketitle
\thispagestyle{plain}
\pagestyle{plain}

\begin{abstract}
Task scheduling is a well-studied problem in the context of optimizing the Quality of Service (QoS) of cloud computing environments. In order to sustain the rapid growth of computational demands, one of the most important QoS metrics for cloud schedulers is the execution cost. In this regard, several data-driven deep neural networks (DNNs) based schedulers have been proposed in recent years to allow scalable and efficient resource management in dynamic workload settings. However, optimal scheduling frequently relies on sophisticated DNNs with high computational needs implying higher execution costs. Further, even in non-stationary environments, sophisticated schedulers might not always be required and we could briefly rely on low-cost schedulers in the interest of cost-efficiency. Therefore, this work aims to solve the non-trivial meta problem of online dynamic selection of a scheduling policy using a surrogate model called MetaNet. Unlike traditional solutions with a fixed scheduling policy, MetaNet on-the-fly chooses a scheduler from a large set of DNN based methods to optimize task scheduling and execution costs in tandem. Compared to state-of-the-art DNN schedulers, this allows for improvement in execution costs, energy consumption, response time and service level agreement violations by up to 11, 43, 8 and 13 percent, respectively. 
\end{abstract}

\section{Introduction}
\label{sec:introduction}

\IEEEPARstart{T}{he} onset of the Artificial Intelligence (AI) and Deep Learning (DL) era has led to a recent shift in computation from hand-encoded algorithms to data-driven solutions~\cite{gill2019transformative}. One such impact of DL is on resource management in distributed computational paradigms~\cite{boudi2021ai, tuli2021mcds}. The most popular such paradigm, cloud computing, harnesses the data processing capacities of multiple devices and provides services at scale with high Quality of Service (QoS). The popularity of cloud architectures is primarily attributed to the ability to provision resources on demand, having significant cost benefits, both for users as well as cloud providers. However, in the age of the Internet of Things (IoT), wherein millions of connected devices produce data that needs to be processed, the demand of computational resources has increased~\cite{cai2016iot}. In such cases, it becomes crucial to curtail the operational costs of cloud machines. This calls for efficient resource management schemes, such as task scheduling policies, to execute workloads on cloud infrastructures within tight cost budgets.

\begin{figure}[t]
    \centering
    \includegraphics[width=0.9\linewidth]{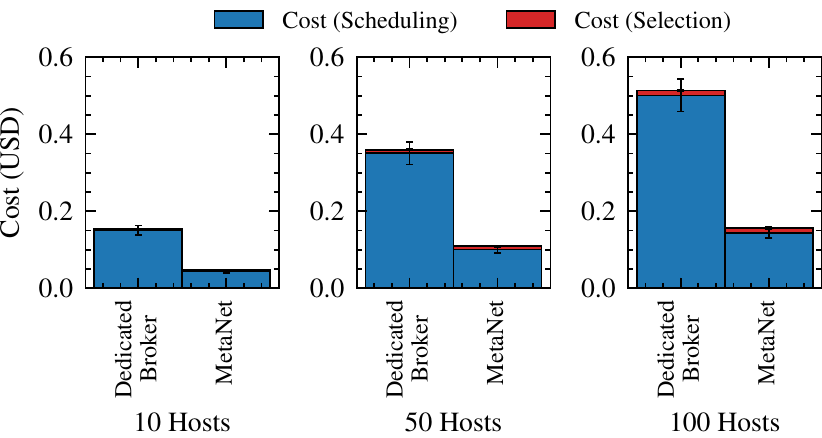}
    \caption{Comparison between cost per task with a dedicated broker node and 10/50/100 cloud hosts as worker nodes. The scheduling policy is run as a serverless function decided by MetaNet. The broker is a cloud VM computational capacity of which depends on the number of worker nodes and the complexity of the DNN based scheduling approach.}
    \label{fig:motivation}
\end{figure}

\textbf{Background and Motivation.} In recent years, the state-of-the-art resource management solutions, which particularly focus on optimal placement of tasks on cloud virtual machines (VMs), leverage data-driven DL methods~\cite{decisionnn, tuli2021hunter, semidirect, tuli2021cosco, graf, tuli2021gosh}. Such methods typically rely on a trace of resource utilization characteristics and optimization metrics of tasks and cloud hosts and are commonly referred to as \textit{trace-driven} schedulers. They utilize such traces to train a deep neural network (DNN) to estimate a set of Quality of Service (QoS) parameters and run optimization strategies to find the best scheduling decision for each incoming task. However, most prior work assumes a broker-worker model, wherein the scheduling policies are periodically run on the broker and the incoming tasks are executed on worker nodes~\cite{tuli2019fogbus}. Here, the schedulers aim to optimize only the amortized task execution costs and typically ignore the cost implications of running the scheduling policies~\cite{alkhanak2015cost, tuli2021pregan}. In such cases, having a dedicated broker is often cost inefficient due to the sparse computational requirement in discrete-time control settings, wherein the task placement decisions are taken at fixed scheduling intervals~\cite{tuli2021cosco}. To tackle this, we resort to paradigms such as Function as a Service (FaaS) that allow execution of DL models as serverless functions, only costing us for the run time of each model. Such stateless function calls work in discrete-time control settings, such as task scheduling, as the decisions are independent~\cite{tuli2021cosco}. In such a case, if we consider each DNN based scheduler as a serverless function and the selection procedure as a task for a worker node, we can work without the broker node altogether, leading to significant cost gains (see Figure~\ref{fig:motivation}). The figure demonstrates that in a case when MetaNet is run on worker nodes and its selected scheduling policies are run as serverless functions, it can lead to up to 68\% lower costs associated with task scheduling. 

\textbf{Challenges.} The problem of selecting a scheduler at each scheduling interval is non-trivial. There is a tradeoff when selecting scheduling policies. Simple schedulers that rely on lightweight DNNs are cost-efficient as they do not impose high computational requirements; however, they do not provide cost-optimal scheduling decisions~\cite{alkhanak2015cost}. On the other hand, schedulers that utilize sophisticated DL models have high execution times, translating to higher costs when running them as serverless functions. However, compared to simpler DNNs, they tend to provide cost-efficient task scheduling decisions. We provide a more evidence based discussion in Section~\ref{sec:experiments}. In non-stationary environments where incoming tasks typically follow bursty workload characteristics~\cite{traverso2013temporal, tuli2022tranad} it may be more cost-efficient to utilize a simpler DNN in trivial cases in lieu of a sophisticated one~\cite{chen2021deep}. This requires dynamic selection of the scheduling policy based on the latest state of the cloud environment. As these decisions need to be taken quickly and considering many resources together, relying on expert-based scheduler selection is infeasible, giving rise to the need for an automated selection approach. 

\begin{figure}[t]
    \centering
    \includegraphics[width=\linewidth]{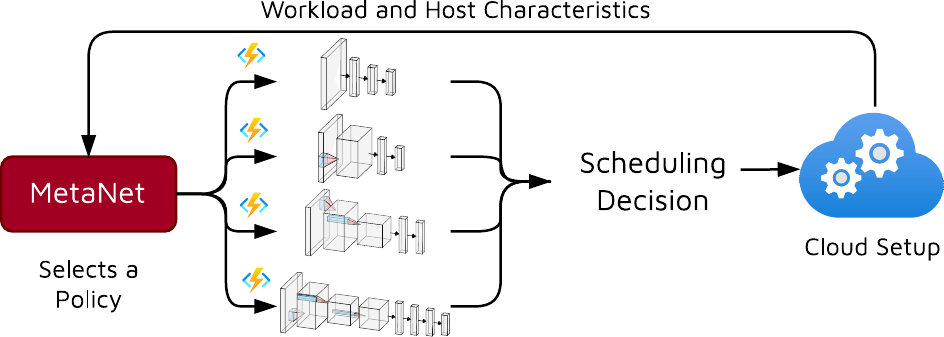}
    \caption{MetaNet Pipeline.}
    \label{fig:overall}
\end{figure}

\textbf{Contributions.} One way to tackle the above problem is to build a tunable scheduler from the ground-up that can make either quick and sub-optimal or slow and optimal scheduling decisions. This tuning parameter that affects its computational requirements can be dynamically updated to ensure cost-efficiency. However, in this work, to leverage the recent scientific advances in task scheduling and build a policy agnostic solution, we choose a set of state-of-the-art methods. This work aims to solve the meta-problem of on-the-fly selection of a scheduling policy for a cloud computing environment wherein the incoming tasks are executed on worker nodes and scheduling decisions are run as serverless functions. To solve this meta-problem, we develop the proposed solution that we call \underline{MetaNet}. A preliminary version of this work was presented as a poster in ACM SIGMETRICS Conference 2022~\cite{tuli2022metanet}. Compared to the previous version, we provide an end-to-end model description with an advanced DNN as surrogate model to predict the task execution costs and scheduling time for each policy. MetaNet selects the most cost-efficient policy at each scheduling interval using such online estimates. An overview of the complete pipeline is presented in Figure~\ref{fig:overall}. As we dynamically update the policy to tradeoff between simple and sophisticated DL schedulers, this facilitates MetaNet to reduce overall operational costs, energy consumption, response time and Service Level Agreement (SLA) violation rates by up to 11\%, 43\%, 8\% and 13\% respectively compared to state-of-the-art schedulers in a static management fashion.  

\section{Related Work}
\label{sec:related_work}

Recent work in scheduling for cloud computing environments has demonstrated that AI-based solutions are not only faster, but can also scale efficiently compared to traditional heuristic and classical optimization techniques~\cite{semidirect, decisionnn, tuli2021cosco, tuli2021hunter, tuli2021gosh}. Most contemporary dynamic resource management methods decouple the decision-making problem into two stages: QoS prediction and decision optimization~\cite{uahs}. This is commonly referred to as the \textit{predict+optimize} framework in literature and is agnostic to the decision type~\cite{semidirect}. We classify the state-of-the-art schedulers as evolutionary and surrogate-based optimization methods.

\textbf{Evolutionary Optimization.} This class of methods forecasts QoS of a future state of a cloud system and needs data corresponding to historical QoS traces of the same system. Several methods have been proposed that leverage a forecasting model. For instance, a class of methods utilizes regression models such as Linear Regression (LR)~\cite{hyndman2018forecasting} or Gaussian Process Regression~\cite{chen2018resource, uahs}. Others utilize auto-regressive models such as AutoARIMA~\cite{arima} based forecasting. Recent works utilize DNNs to perform forecasting, for instance, using LSTM neural networks~\cite{lstm} (we include AutoARIMA for completeness, albeit it not being a DL-based solution). Using a QoS prediction model, several previous works optimize the provisioning decision to minimize execution costs or maximize the utilization ratio. Conventional methods often use evolutionary search strategies such as Ant Colony Optimization (ACO)~\cite{aco}, which has been shown to exhibit state-of-the-art QoS scores in recent work, particularly due to the property of quickly converging to optima~\cite{cahs}.

\textbf{Surrogate Optimization.} However, even with the advantages of scalability and quick convergence to optima, few prior works use gradient-based methods as neural approximators are not consistent with the convexity/concavity requirements of such methods~\cite{nandi2001artificial, tuli2022carol}. A method that resolves this is GRAF~\cite{graf} that uses a graph neural network (GNN) as a surrogate model to predict service latencies and operational costs for a given scheduling decision and uses gradient-based optimization to minimize the service latencies or execution costs. To do this, it uses the concept of neural network inversion~\cite{nninversion}, wherein the method evaluates gradients of the objective function with respect to inputs and runs optimization in the input space. Other methods, such as Decision-NN, combine the prediction and optimization steps by modifying the loss function to train neural networks in conjunction with the optimization algorithm~\cite{decisionnn}. This method uses a neural network as a surrogate model to directly predict optimization objectives such as costs to run decision optimization. However, simple gradient-based approaches tend to often get stuck in local optima. This problem is alleviated by momentum and annealing in schedulers like GOBI and GOSH~\cite{tuli2021cosco, tuli2021gosh}. Such methods take the scheduling decision and state of the cloud system as resource utilization characteristics of workloads and cloud nodes and output a QoS estimate such as execution cost or scheduling time. GOSH, unlike GOBI, uses a stochastic estimate of the QoS and higher-order gradient updates for more accurate QoS predictions and faster convergence compared to first-order optimization strategies.

It is known that the gradient-free optimization methods are computationally efficient, but provide poorer QoS results. Gradient-based methods provide improved QoS scores at the cost of higher compute requirements. For a broad coverage, we consider methods from both classes discussed above. The methods: ARIMA+ACO, LSTM+ACO, DecisionNN, Semi-Direct, GRAF, GOBI and GOSH have been shown to give state-of-the-art performance and hence used as baselines in our experiments to validate the efficacy of MetaNet.

\section{Methodology}
\label{sec:method}

\begin{figure}
    \centering
    \includegraphics[width=0.8\linewidth]{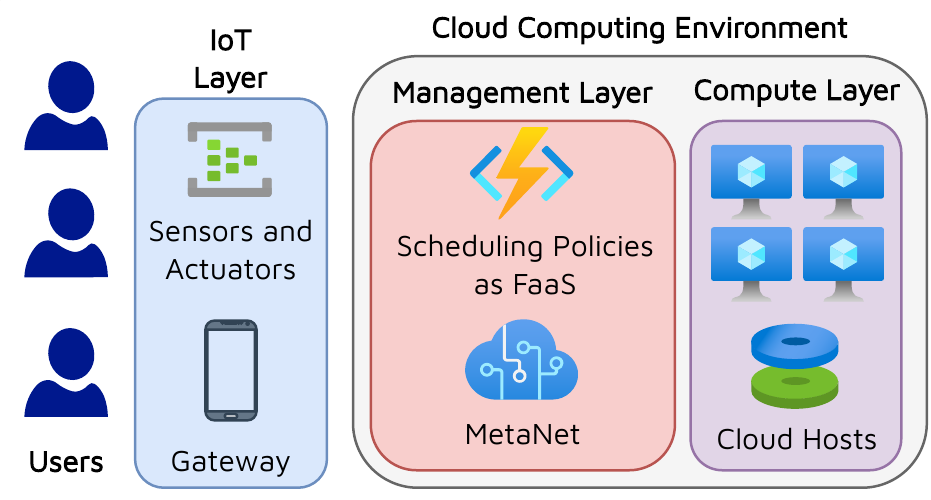}
    \caption{System Model.}
    \label{fig:system}
\end{figure}

\subsection{System Model and Problem Formulation}
In this work, we target a standard heterogeneous cloud computing environment where all nodes are in the same Wide Area Network (WAN); see Figure~\ref{fig:system} for an overview. Our cloud computing environment consists of multiple VMs that process a set of independent workloads. Tasks are container instances that need to be processed, generated from the sensors and actuators in the IoT layer and communicated to the compute nodes via gateway devices. We do not have any cloud broker; instead, we run the scheduling policies as FaaS functions on a serverless platform such as AWS Lambda or Azure Serverless. As is common in prior work~\cite{tuli2021cosco, basu2019learn, tuli2021hunter}, we consider a discrete-time control problem, \textit{i.e.}, we divide the timeline into fixed-size execution intervals (of $\Delta$ seconds) and denote the $t$-th interval by $I_t$. A scheduling decision is made at the start of each interval for all incoming tasks in the previous interval. We consider a bounded execution with $T$ intervals; thus, $t \in \{1, \ldots, T\}$. The decision of which scheduling policy to execute as well as the scheduling decision is made at the start of each interval. We consider a set of scheduling policies $\mathcal{P}$ of size $q$. A summary of notation is presented in Table~\ref{tab:symbols}.

We assume that there are a $m$ number of host machines in the cloud compute layer and denote them as $\mathcal{H}$. We also consider that there are $n_t$ workloads in the system in interval $I_t$ and denote the set of workloads as $\mathcal{W}_t$. The feature vector of task $w^i_t \in \mathcal{W}_t$ ($i$ denoted the index of a task) includes the workload information with CPU utilization in terms of the number of instructions per second (IPS), denoted by $c^i_t$; RAM utilization in GBs, denoted by $r^i_t$; and disk storage utilization in GBs, denoted by $s^i_t$. Here, $k \in \{1, \ldots, n_t\}$. Thus, the feature vector for task $w^i_t$ is denoted by $W^i_t = [c^i_t, r^i_t, s^i_t]$. The collection of feature vectors of all active tasks in $I_t$ is denoted by $W_t$. We similarly define the feature vector of host $h^j \in \mathcal{H}$ ($j$ denoted the index of a host) at interval $I_t$ as $H^j_t = [\bar{c}^j_t, \bar{r}^j_t, \bar{s}^j_t]$ where $\bar{c}^j_t, \bar{r}^j_t, \bar{s}^j_t$ denote the CPU utilization in terms of IPS, RAM utilization in GBs and disk storage utilization in GBs averaged over interval $I_t$. The collection of feature vectors of all hosts at interval $I_t$ is denoted by $H_t$. We also form a graph denoted by $S_t = (V^S_t, E^S_t)$, which is a bi-partite graph with nodes of two types: tasks and hosts. The edges of the graph $(w^i_t, h^j) \in E^S_t$ correspond to the allocation decision where task $w^i_t$ is allocated to host $h^j$. Each task and host has a feature vector corresponding to the IPS, RAM, and disk storage consumption as before. 

\begin{table}[t]
    \centering
    \caption{Table of Notations}
    \label{tab:symbols}
    \resizebox{\linewidth}{!}{
    \begin{tabular}{@{}ll@{}}
    \toprule 
    Symbol & Meaning\tabularnewline
    \midrule
    $I_t$ & $t$-th interval \tabularnewline
    $\mathcal{P}$ & Set of scheduling policies of size $q$ \tabularnewline
    $\mathcal{H}$ & Set of available hosts of size $m$ \tabularnewline
    $\mathcal{W}_t$ & Set of tasks in $I_t$ of size $n_t$ \tabularnewline
    $H^j_t$ & Feature vector of host $h^j \in \mathcal{H}$ at $I_t$ \tabularnewline
    $W^i_t$ & Feature vector of task $w^i_t \in \mathcal{W}_t$ \tabularnewline
    $S_t$ & Scheduling decision in $I_t$ as a bi-partite graph \tabularnewline
    $\phi^k_t$ & Amortized task execution cost for scheduler $p^k \in \mathcal{P}$ in $I_t$ \tabularnewline
    $\omega^k_t$ & Scheduling time for scheduler $p^k \in \mathcal{P}$ in $I_t$ \tabularnewline
    \bottomrule
    \end{tabular}}
\end{table}

If we use the scheduling policy $p^k \in \mathcal{P}$ at interval $I_t$, then we denote the scheduling decision by $p^k(t)$. Here $q \in \{1, \ldots, q\}$. At interval $I_t$ we denote the tasks that complete in this interval by $\eta_t \subseteq \mathcal{W}_t$. We denote the cost per second for host $h^j$ as $\mu^j$. Thus, for interval $I_t$ we get amortized task execution cost for a scheduler $p^k$ as
\begin{equation}
\label{eq:cost}
    \phi^k = \frac{\sum_{h^j \in \mathcal{H}} \mu^j \cdot \Delta}{|\eta_t|}, \text{ s.t. } S_t = p^k(t).
\end{equation}
The collection of all execution costs is denoted by $\phi$. Similarly, we denote the scheduling time for scheduler $p^k$ at interval $I_t$ by $\omega^k_t$. For simplicity, we consider the cost for running a serverless function for a unit second as a static parameter and denote it by $\rho$. Thus, for interval $I_t$ we get the scheduling cost for a scheduler $p^k$ by
\begin{equation}
    \omega^k_t \cdot \rho.
\end{equation}

\begin{figure*}
    \centering
    \includegraphics[width=0.9\linewidth]{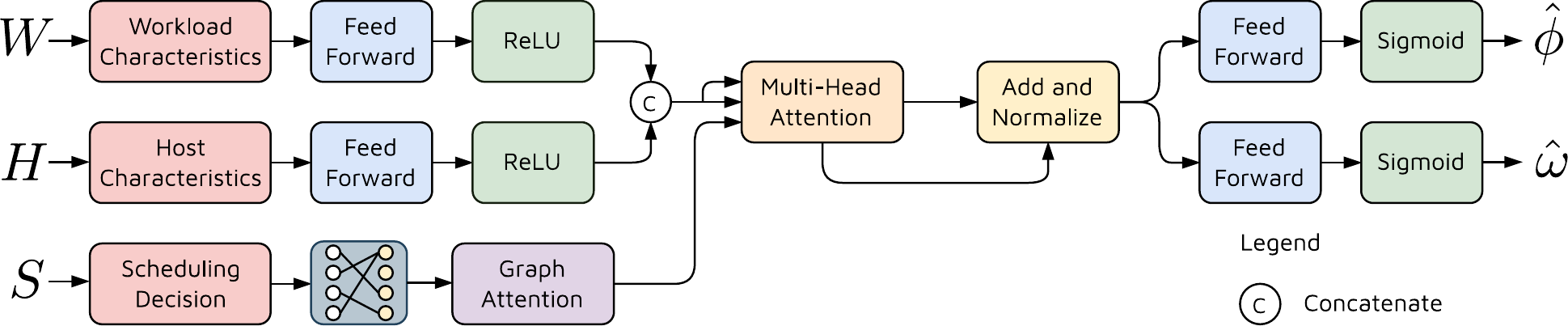}
    \caption{MetaNet Surrogate Model.}
    \label{fig:model}
\end{figure*}

In this work, we create a surrogate model $f_\theta$ that is a DNN with parameters $\theta$. Given a system state at the start of interval $I_t$ characterized by $[W_{t-1}, H_{t-1}, S_{t-1}]$, $f_\theta$ estimates the average task execution cost on the worker nodes for each scheduler $p^k \in \mathcal{P}$ as $\hat{\phi}^k_t$ and the scheduling time of each scheduler $p^k$ as $\hat{\omega}^k_t$.  The collections of $\hat{\phi}^k$ and $\hat{\omega}^k$ for all $k$ in $I_t$ are denoted by $\hat{\phi}_t$ and $\hat{\omega}_t$, respectively. In such a case, our problem can be formulated as
\begin{equation}
\label{eq:problem}
\begin{aligned}
& \underset{\theta}{\text{minimize}} 
& & \sum_{t=1}^T \phi^\pi_t + \omega^\pi_t \cdot \rho \\
& \text{subject to}
& & S_t = p^k(t), \forall\ t \\
&&& \pi = \argmin_k ( \hat{\phi}^k_t + \hat{\omega}^k_t \cdot \rho ),  \forall\ t  \\
&&& \hat{\phi}_t, \hat{\omega}_t = f_\theta(W_{t-1}, H_{t-1}, S_{t-1}), \forall\ t,
\end{aligned}
\end{equation}
where $f_\theta$ is a surrogate of execution cost and scheduling time of the cloud environment. Predicting the cost for each interval is crucial as it is a variable that depends on the load characteristics of the hosts in the cloud setup as well as the running workloads. Predicting scheduling time is crucial as it is not a static metric for many DL based schedulers. DL-based schedulers have sophisticated search strategies such as ACO and gradient-based optimization. Each has a different convergence criterion that changes with the number of workloads and volatility in the resource utilization characteristics.

\subsection{MetaNet Model}

Considering the above formulation, instead of optimizing the weights of the neural network $f_\theta$ using the execution style as denoted by~\eqref{eq:problem}, to simplify training, we optimize $\theta$ such that the predicted $\bar{\phi}^k_t, \bar{\omega}^k_t$ match the ground-truth values $\phi^k_t, \omega^k_t$ for a scheduler $p^k$ in interval $I_t$. We do this as the miniizer of $k$ over $(\hat{\phi}^k_t + \hat{\omega}^k_t \cdot \rho)$ would match that over $(\phi^k_t + \omega^k_t \cdot \rho)$. Thus, we train $f_\theta$ to act as a surrogate of the $\phi$ and $\omega$ metrics. We call this surrogate as the MetaNet model, an overview of which is given in Figure~\ref{fig:model}. For the sake of simplicity and without loss of generality, from now we drop the $t$ index whenever this is not ambiguous and use $W$, $H$, $S$, $\phi^k$, $\omega^k$, $\bar{\phi}^k$ and $\bar{\omega}^k$ instead.

We now describe the complete pipeline in detail. The input workload and host characteristics $W$ and $H$ are first converted to matrices of size $m \times 3$ and $n \times 3$. Here, the $3$ features correspond to the CPU, RAM and storage characteristics. The scheduling decision $S$ is encoded as a graph with each node feature vector of size $3$. We use a composite neural network as MetaNet and infer over the inputs as we describe next.

To infer over the workload and host characteristics, $W$ and $H$, we utilize a fully-connected network (FCN), also referred to as feed-forward neural network in literature. We use $\mathrm{ReLU}$ activation function with the FCN to generate
\begin{align}
\begin{split}
    E^W = \mathrm{ReLU}(\mathrm{FeedForward}(W)),\\
    E^H = \mathrm{ReLU}(\mathrm{FeedForward}(H)).
\end{split}
\end{align}
To infer over the scheduling decision graph $S$, we use a graph-attention network (GAT) network~\cite{gat}. Graph attention operation performs convolution operation for each node over its neighbors and uses dot product self-attention to aggregate feature vectors. To build an approach agnostic to the number of task or hosts, we create a new global node connected to each workload and host node~\cite{xie2020mgat}. This gives the graph attention operation as
\begin{equation}
    E^S = \mathrm{Sigmoid} \bigg( \frac{1}{n} \sum_{i=1}^n \theta^W_{\rm GAT} W^i + \frac{1}{m} \sum_{j=1}^m \theta^H_{\rm GAT} H^j \bigg),
\end{equation}
where $\theta^H_{\rm GAT}$ and $\theta^W_{\rm GAT}$ are the weight matrices for the GAT network. Now that we get the embedding outputs $E^W$, $E^H$ and $E^S$, we apply multi-headed attention~\cite{vaswani2017attention}. For any three input tensors $Q$, $K$ and $V$, we define multi-head self attention~\cite{vaswani2017attention} as passing it through $z$ (number of heads) feed-forward layers to get $Q_i$, $K_i$ and $V_i$ for $i \in \{1, \ldots, z\}$, and then applying attention as
\begin{align}
\begin{split}
    \mathrm{MultiHeadAtt}(Q, K, V) &= \mathrm{Concat}(X_1, \ldots, X_z)\\
     X_i &= \mathrm{Attention}(Q_i, K_i, V_i).
\end{split}
\end{align}
Multi-Head Attention allows the model to jointly attend to information from different representation sub-spaces at different positions. We first concatenate the two embeddings to generate
\begin{equation}
    E^{W, H} = [E^W, W^H].
\end{equation}
Then, we perform the operation
\begin{align}
\begin{split}
\label{eq:encoder}
    E^M &= \mathrm{MultiHeadAtt}(E^{W, H}, E^{W, H}, E^S),\\
    E &= \mathrm{LayerNorm}(E^{W, H} + E^M),
\end{split}
\end{align}
where the $\mathrm{LayerNorm}$ operation normalizes the output for stable training. The intuition behind using the self-attention module is to focus on key task and host characteristics that affect the scheduling time and execution costs. We finally generate the scheduling time and execution cost estimates using feed-forward networks as
\begin{align}
\begin{split}
\label{eq:encoder}
    \hat{\phi} &= \mathrm{Sigmoid}(\mathrm{FeedForward}(E)),\\
    \hat{\omega} &= \mathrm{Sigmoid}(\mathrm{FeedForward}(E)).\\
\end{split}
\end{align}
Here, $\hat{\phi}$ and $\hat{\omega}$ consists of estimates $\hat{\phi}^k, \hat{\omega}^k$ for each scheduling policy $p^k \in \mathcal{P}$. As we keep a static size set $\mathcal{P}$, $\hat{\phi}, \hat{\omega}$ are vector-like outputs. The $\mathrm{Sigmoid}$ operation gives an estimate in the normalized form, belonging to the range $(0, 1)$~\cite{karlik2011performance}. Thus, for any input $(W, H, S)$, the complete neural network can be described as
\begin{equation}
    \hat{\phi}, \hat{\omega} = f_\theta(W, H, S)
\end{equation}

\subsection{Offline Training of MetaNet}
\label{sec:training}

To train the above described MetaNet neural network $f_\theta$, we collect traces from a cloud computing environment. To collect data for training, we execute the scheduling policies $\mathcal{P}$, each for $\Gamma$ scheduling intervals, and collect a dataset as a collection of tuples
\begin{equation}
    \label{eq:dataset}
    \Lambda = \{(k, W_{t-1}, H_{t-1}, S_{t-1}, \phi^k_t, \omega^k_t)\}_{t = 1}^{\Gamma \cdot q},
\end{equation}
where $q$ is the number of scheduling policies in $\mathcal{P}$. We initialize $W_0$ as $H_0$ zero-matrices and $S_0$ as an empty graph.  As the cost and scheduling time in the dataset are not in the range $(0, 1)$ as in the output of the neural network, we also find the maximum of these two variables as
\begin{align}
\begin{split}
\label{eq:denorm}
    \phi^k_{max} &= \max_{t = 1}^{\Gamma \cdot q} \phi^k_t,\\
    \omega^k_{max} &= \max_{t = 1}^{\Gamma \cdot q} \omega^k_t.
\end{split}
\end{align}
This allows us to denormalize the neural network output and bring it to the same range as the one in the dataset. However, this is sensitive to the outliers in the data. We thus perform outlier removal using the Local Outlier Factor (LOF) approach~\cite{alghushairy2020review} prior to obtaining these denormalizaing coefficients.

\begin{figure}[t]
    \centering
    \includegraphics[width=0.9\linewidth]{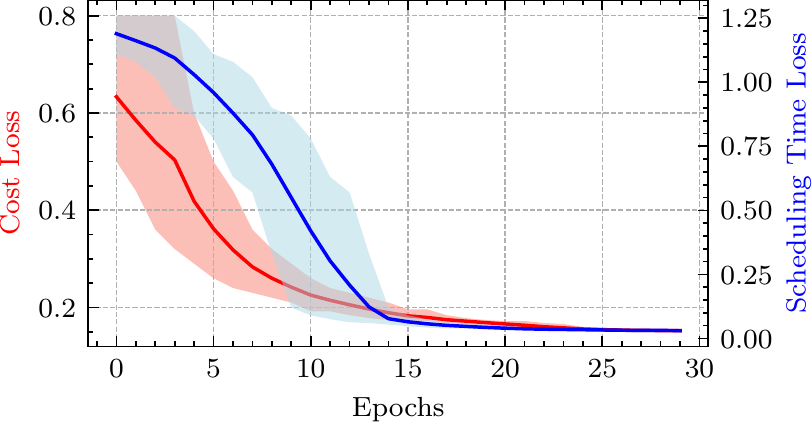}
    \caption{Training plots for the MetaNet model on the validation set. The model converges in 30 training epochs.}
    \label{fig:convergence}
\end{figure}

Considering we now have a dataset $\Lambda$ collected as shown in~\eqref{eq:dataset}, for each datapoint $(k, W, H, S, \phi^k, \omega^k)$, we define two loss functions: cost loss $L^C$ and scheduling time loss $L^S$ as
\begin{align}
\begin{split}
\label{eq:loss}
    L^C &= \norm{\phi^k - \phi^k_{max} \cdot \hat{\phi}^k}_{2},\\
    L^S &= \norm{\omega^k - \omega^k_{max} \cdot \hat{\omega}^k}_{2},
\end{split}
\end{align}
where $\hat{\phi}, \hat{\omega} = f_\theta(W, H, T)$. The cost loss, $L^C$, is an L2-norm between the ground-truth execution cost and the denormalized predicted cost for scheduler $p^k$ as $\phi^k_{max} \cdot \hat{\phi}^k$. Similarly, the scheduling time cost, $L^S$, is the L2-norm between ground-truth scheduling time and predicted scheduling time $\omega^k_{max} \cdot \hat{\omega}^k$.

To train the MetaNet neural model, we use the cumulative loss as the sum of the cost loss and scheduling time loss as
\begin{equation}
    L = L^C + L^C,
\end{equation}
to train the model. We use the AdamW optimizer~\cite{saleh2019dynamic} with a learning rate $0.005$ and randomly sample 80\% of data to get the training set and the rest as the validation set. We use a weight regularization parameter as $10^{-5}$ to avoid overfitting~\cite{han2015learning}. All model training is performed on a separate server with this configuration: Intel Xeon Silver 2.20-3.00 GHz CPU, 16GB RAM, Nvidia RTX 3090 and Ubuntu 18.04 LTS OS. We use the early-stopping as the convergence criterion wherein we stop the training as soon as we see the validation loss increase. Using this procedure, we get a trained neural network $f_\theta$ that acts as a surrogate of the execution cost and scheduling time of all policies in $\mathcal{P}$. The trends of cost loss and scheduling time loss for the dataset collected from policies and setup described in Section~\ref{sec:experiments} are shown in Figure~\ref{fig:convergence}. The plot shows that the model converges in 30 epochs for this setup. For a new setup with a different set of policies, the model would need to be retrained or fine-tuned using one of the transfer learning methods~\cite{pan2009survey}. 

\begin{algorithm}[t]
    \begin{algorithmic}[1]
    \Require
    \Statex Pre-trained surrogate model $f_\theta$
    \Statex Dataset used for training $\Lambda$
    \Statex Set of schedulers $\mathcal{P}$
    \Procedure{$\text{MetaNet}$}{scheduling interval $I_t$}
        \State \textbf{if} (t == 0)
        \State \hspace{\algorithmicindent} Initialize matrices $H_0, W_0$ as zero
        \State \hspace{\algorithmicindent} Initialize $S_0$ as empty graph
        \State \hspace{\algorithmicindent} Initialize $\phi^k_{max}, \omega^k_{max}$ as per~\eqref{eq:denorm}
        \State Get $W_{t-1}, H_{t-1}, S_{t-1}$ \label{line:getmetrics}
        \State Select host $h^\kappa$ as $\kappa = \argmin_j \bar{c}^j_{t-1}$ to run MetaNet \label{line:selecthost}
        \State $\hat{\phi}_t, \hat{\omega}_t \gets f_\theta(W_{t-1}, H_{t-1}, S_{t-1})$ \label{line:predict}
        \State $\pi = \argmin_k \phi^k_{max} \cdot \hat{\phi}^k_t =  \omega^k_{max} \cdot \hat{\omega}^k_t$ \label{line:sched}
        \State $S_t = p^\pi(t)$ \label{line:sched2}
        \State Schedule tasks using $S_t$
        \State Get $W_t, H_t, S_t, \phi^\pi_t, \omega^\pi_t$
        \State Fine-tune $f_\theta$ with loss \label{line:tune}
        \Statex \quad \quad $L = \norm{\phi^\pi - \phi^\pi_{max} \cdot \hat{\phi}^\pi}_{2} + \norm{\omega^\pi - \omega^\pi_{max} \cdot \hat{\omega}^\pi}_{2},$
        \Statex \quad \quad where $\hat{\phi}, \hat{\omega} \gets f_\theta(W_t, H_t, S_t) $
        \State \textbf{return} $S_t$
    \EndProcedure
    \end{algorithmic}
\caption{The MetaNet scheduler}
\label{alg:metanet}
\end{algorithm}

\begin{table*}[]
    \centering
    \caption{Host characteristics of Microsoft Azure distributed cloud environment.}
    \resizebox{\textwidth}{!}{
        \begin{tabular}{@{}lcccccccccc@{}}
    \toprule 
    \multirow{2}{*}{Name} & \multirow{2}{*}{Quantity} & Core & \multirow{2}{*}{MIPS} & \multirow{2}{*}{RAM} & RAM & Ping & Network & Disk & Cost & \multirow{2}{*}{Location}\tabularnewline
     &  & count &  &  & Bandwidth & time & Bandwidth & Bandwidth & Model & \tabularnewline
    \midrule
    \multicolumn{11}{c}{\textbf{Private Cloud Layer}}\tabularnewline
    \midrule
    Azure B2s server & 4 / 20 / 40 & 2 & 4029 & 4295 MB & 372 MB/s & 3 ms & 1000 MB/s & 13.4 MB/s & 0.0472 \$/hr & London, UK\tabularnewline

    Azure B4ms server & 2 / 10 / 20 & 4 & 8102 & 17180 MB & 360 MB/s & 3 ms & 1000 MB/s & 10.3 MB/s & 0.1890 \$/hr & London, UK\tabularnewline
    \midrule
    \multicolumn{11}{c}{\textbf{Public Cloud Layer}}\tabularnewline
    \midrule
    Azure B4ms server & 2 / 10 / 20 & 4 & 8102 & 17180 MB & 360 MB/s & 76 ms & 1000 MB/s & 10.3 MB/s & 0.166 \$/hr & Virginia, USA\tabularnewline

    Azure B8ms server & 2 / 10 / 20 & 8 & 2000 & 34360 MB & 376 MB/s & 76 ms & 2500 MB/s & 11.64 MB/s & 0.333 \$/hr & Virginia, USA\tabularnewline
    \bottomrule 
    \end{tabular}}
    \label{tab:hosts}
\end{table*}

\subsection{Dynamic Policy Selection Using MetaNet}

Using a surrogate model $f_\theta$ that has been trained offline using the dataset as described in Section~\ref{sec:training}, we now describe how to dynamically select scheduling policies to optimize the total cost of a cloud computing platform. A summary is given in Algorithm~\ref{alg:metanet}. We first store the saved surrogate model $f_\theta$ into a central network-attached-storage (NAS) that is accessible to all worker nodes. At start the scheduling interval $I_t$, we get the workload and host characteristics $W_{t-1}, H_{t-1}$ with the scheduling decision $S_{t-1}$ (line~\ref{line:getmetrics} in Alg.~\ref{alg:metanet}). In our setup, as we do not have any broker node, we run the MetaNet model in the worker node with the least CPU utilization (line~\ref{line:selecthost} in Alg.~\ref{alg:metanet}). On this worker node we use the surrogate model to predict cost and scheduling time (line~\ref{line:predict}) as
\begin{equation}
    \label{eq:predict}
    \hat{\phi}_t, \hat{\omega}_t = f_\theta(W_{t-1}, H_{t-1}, S_{t-1}).
\end{equation}
The scheduling policy is then decided (line~\ref{line:sched}) as 
\begin{equation}
    \label{eq:decide}
    p^\pi, \text{ s.t. } \pi = \argmin_k \phi^k_{max} \cdot \hat{\phi}^k_t + \omega^k_{max} \cdot \hat{\omega}^k_t.
\end{equation}
With the decided scheduling policy $p^\pi$, the scheduling decision for interval $I_t$ is then $S_t = p^\pi(t)$ (line~\ref{line:sched2}). 
With the execution of scheduling decision $S_t$, we can now also generate another datapoint $(\pi, W_t, H_t, S_t, \phi^\pi_t, \omega^\pi_t)$ as per the running cost ($\omega^\pi_t$) and scheduling time ($\phi^\pi_t$). We can utilize this datapoint to fine-tune the MetaNet surrogate (line~\ref{line:tune}) using the loss function
\begin{equation}
    \label{eq:tuneloss}
    L = \norm{\phi^\pi - \phi^\pi_{max} \cdot \hat{\phi}^\pi}_{2} + \norm{\omega^\pi - \omega^\pi_{max} \cdot \hat{\omega}^\pi}_{2}, 
\end{equation}
similar to the loss function described~\eqref{eq:loss}. This allows us to adapt the model with changing workload or host characteristics in the environment. The volatility in the environment can be present due to mobility of the users or dynamism in the workload characteristics. Periodic model fine-tuning facilitates optimum performance of the surrogate model even in when the environment characteristics change with time.

Overall, MetaNet selects a scheduling policy on-the-fly as per the system states. To do this, at the start of each scheduling interval, it predicts the task execution cost and scheduling time of each policy. It then selects the one with the minimum cost estimate. 

\section{Experiments}
\label{sec:experiments}

\subsection{Implementation and Model Training}
\label{sec:implementation}
To implement MetaNet,  we build upon the execution primitives provided by the COSCO framework~\cite{tuli2021cosco} by modifying and integrating with custom resource provisioning methods, specifically for dynamic selection of the scheduling policies. Further, we use HTTP RESTful APIs for communication and seamless integration of a \texttt{Flask} based web environment to deploy and monitor the resource utilization characteristics of running workloads in our distributed cloud setup~\cite{grinberg2018flask}. Our tasks are executed using Docker containers. We use the Checkpoint/Restore In Userspace (CRIU)~\cite{venkatesh2019fast} tool for container migration when  scheduling decision of a task in terms of the host it is to be placed on for any interval is different from the previous interval. All sharing of resource utilization characteristics across worker nodes to execute the  uses the \texttt{rsync}\footnote{RSync: \url{https://linux.die.net/man/1/rsync}.} utility. We extend the \texttt{Framework} class to allow decentralized decision making. We utilize the HTTP Notification API to synchronize outputs and execute workloads. All schedulers are converted to serverless functions and deployed on Azure Serverless Platform using the RADON framework~\cite{casale2020radon}. We collect the dataset $\Lambda$ by executing all baselines described in Section~\ref{sec:baselines} for $\Gamma = 100$ intervals, with the neural network training time of 17 minutes.  Similarly, we execute all approaches for $T = 1000$ scheduling intervals to generate QoS scores, with each interval being $\Delta = $ 10 seconds long, giving a total experiment time of nearly 2 hours 46 minutes.

\subsection{Setup}
\label{sec:setup}
We consider the complete set of hosts $\mathcal{H}$ to be static with time as is common in prior work for a fixed cloud platform~\cite{uahs, cahs, semidirect}. We use diverse VM types, in our cloud infrastructure, \textit{i.e.}, \texttt{B2s} with a dual-core CPU and 4GB RAM, \texttt{B4ms} with a quad-core CPU and 16GB RAM and \texttt{B8ms} with an octa-core CPU and 32 GB RAM. We consider a geographically distributed cloud environment with 10, 50 and 100 VMs to test the efficacy of the approaches at different scales. A summary is presented in Table~\ref{tab:hosts}. Our environment consists of up to 60 VMs in the \texttt{UK-South} Azure datacenter and up to 40 in the \texttt{East-US} datacenter. This geographically distributed and heterogeneous environment is chosen as per prior work~\cite{tuli2021cosco, basu2019learn, tuli2019edgelens}. To save on costs, we define a host to be active when the CPU utilization is $> 0\%$. We use the \texttt{Azure Automation}\footnote{Azure Automation Service: \url{https://azure.microsoft.com/en-us/services/automation/\#overview}.} service to auto-hibernate and resume VMs based on the CPU utilization of the VMs.

The maximum possible IPS of host machines were set as per the resultant IPS from running the \texttt{sysbench}\footnote{Sysbench: \url{http://manpages.ubuntu.com/manpages/trusty/man1/sysbench.1.html}.} CPU benchmarking tool, RAM and Disk capacities as per the Azure cloud VM specifications. We calculate the IPS, RAM and Disk utilizations, \textit{i.e.}, $c^i_t, r^i_t, s^i_t$ for task $w^i_t$ and $\bar{c}^j_t, \bar{r}^j_t, \bar{s}^j_t$ for host $h^j$ using the \texttt{Docker Inspect}\footnote{Docker SDK: \url{https://docker-py.readthedocs.io/en/stable/client.html}.} utility in Python and the \texttt{iozone}\footnote{IOZone: \url{https://linux.die.net/man/1/iozone}.} linux benchmarking tool. The costs $\mu^j$ are taken from Azure VM pricing calculator\footnote{Azure VM Pricing Calculator: \url{https://azure.microsoft.com/en-gb/pricing/calculator/}.} for the \texttt{UK-South} and \texttt{East-US} Azure datacenters. The serverless Azure cost, $\rho$, is taken from the Azure Serverless pricing calculator\footnote{Azure Serverless Pricing Calculator: \url{https://azure.microsoft.com/en-gb/pricing/details/functions/}.}. The power consumption values of increments of 10\% CPU utilization of Azure VM types are taken from Standard Performance Evaluation Corporation (SPEC) benchmark repository\footnote{SPEC power consumption repository: \url{https://www.spec.org/cloud\_iaas2018/results/}}.

\subsection{Workloads}
\label{sec:workloads}

To generate the tasks in our system, we use the \textit{AIoTBench} applications~\cite{luo2018aiot}. AIoTBench is an AI based cloud computing benchmark suite that consists of various real-world computer vision application instances. The seven specific application types correspond to the neural networks they utilize. These include three typical heavy-weight networks: ResNet18, ResNet34, ResNext32x4d, as well as four light-weight networks: SqueezeNet, GoogleNet, MobileNetV2, MnasNet.\footnote{AIoTBench: \url{https://www.benchcouncil.org/aibench/aiotbench/index.html}. Accessed: 10 May 2022.} This benchmark has been used in our experiments due to its volatile utilization characteristics and heterogeneous resource requirements. The benchmark consists of 50,000 images from the COCO dataset as workloads~\cite{coco}. To evaluate the proposed method in a controlled environment, we abstract out the users and IoT layers described in Section~\ref{sec:method} and use a discrete probability distribution to realize tasks as container instances. Thus, at the start of each scheduling interval, we create new tasks from a Poisson distribution with rate $\lambda = 1.2$, sampled uniformly from the seven applications. The Poisson distribution is a natural choice for a bag-of-tasks workload model, common in edge environments~\cite{mao2016dynamic, basu2019learn}. We run all experiments for 100 scheduling intervals, with each interval being 300 seconds long, giving a total experiment time of 8 hours 20 minutes. We average over five runs and use diverse workload types to ensure the statistical significance of our experiments.

Similar to a non-stationary workload setup where the users sending tasks are mobile, we utilize a mobility model for sensors/actuators in our setup. To factor in the mobility of the users that send the tasks to the cloud setup, we use the \texttt{NetLimiter} tool to tweak the communication latency with the broker node using the mobility model described in~\cite{gilly2020modelling}. Specifically, we use the latency and bandwidth parameters of hosts from the traces generated using the Simulation of Urban Mobility (SUMO) tool~\cite{krajzewicz2012recent} that emulates mobile vehicles in a city like environment. SUMO gives us the parameters like \texttt{ping} time and network bandwidth to simulate in our testbed using \texttt{NetLimiter}.

\begin{figure}
    \centering
    \includegraphics[width=\linewidth]{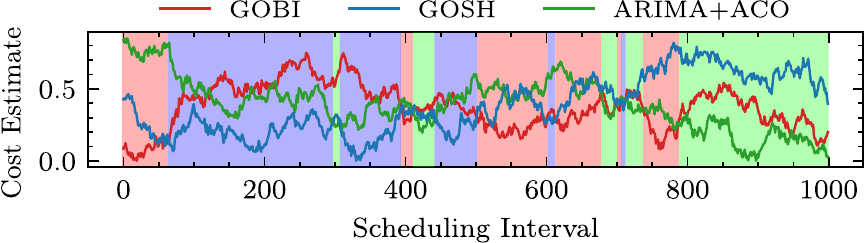}
    \caption{Visualization of predicted costs (line plots) and dynamic scheduler selection (background color) for top three schedulers in MetaNet.}
    \label{fig:vis}
\end{figure}

\begin{figure*}[!t]
    \centering  \setlength{\belowcaptionskip}{-10pt}
    \includegraphics[width=\linewidth]{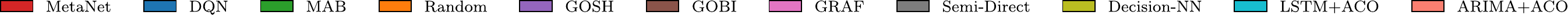} \\
    \includegraphics[width=0.4\linewidth]{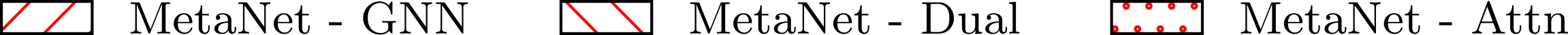} \\
    \subfigure[Energy Consumption]{
    \includegraphics[height=.185\textwidth]{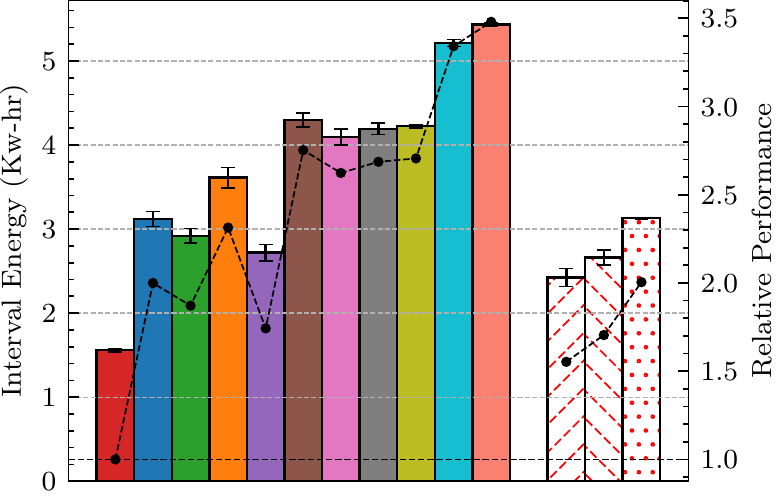}
    \label{fig:energy}
    }
    \subfigure[Response Time]{
    \includegraphics[height=.185\textwidth]{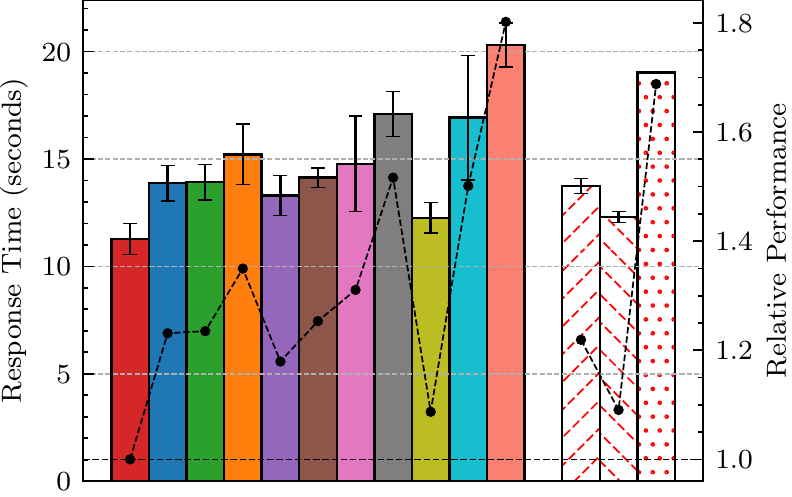}
    \label{fig:rt}
    }
    \subfigure[Waiting Time]{
    \includegraphics[height=.185\textwidth]{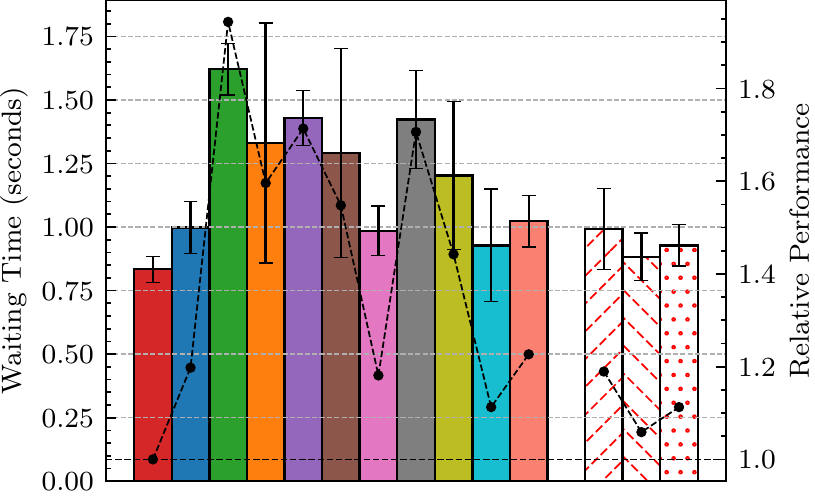}
    \label{fig:wait}
    }\\
    \subfigure[SLA Violation Rate]{
    \includegraphics[height=.185\textwidth]{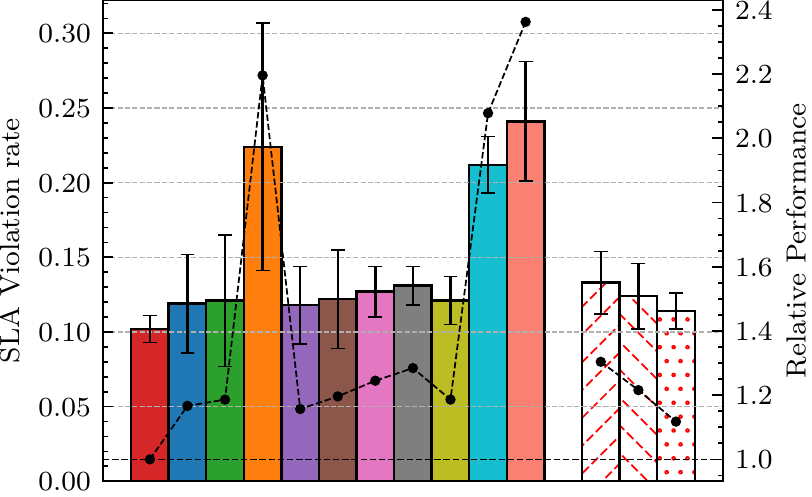}
    \label{fig:sla}
    }
    \subfigure[Jain's Fairness Index]{
    \includegraphics[height=.185\textwidth]{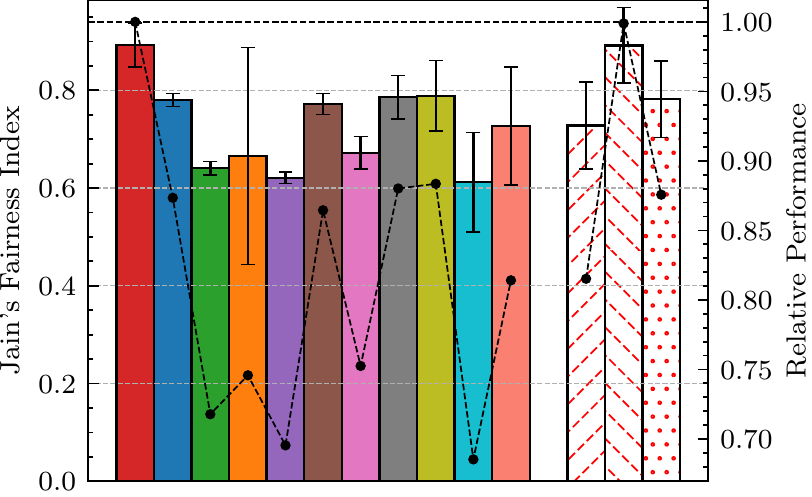}
    \label{fig:fairness}
    }
    \subfigure[CPU Utilization]{
    \includegraphics[height=.185\textwidth]{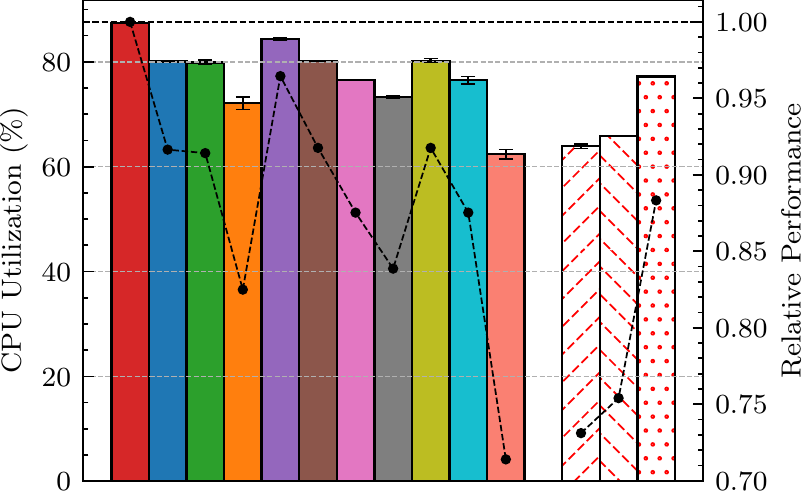}
    \label{fig:cpu}
    }
    \caption{Comparison of QoS parameters (averaged over intervals) of MetaNet against baselines and ablated models.}
    \label{fig:results}
\end{figure*}

\begin{figure}
    \centering
    \includegraphics[width=\linewidth]{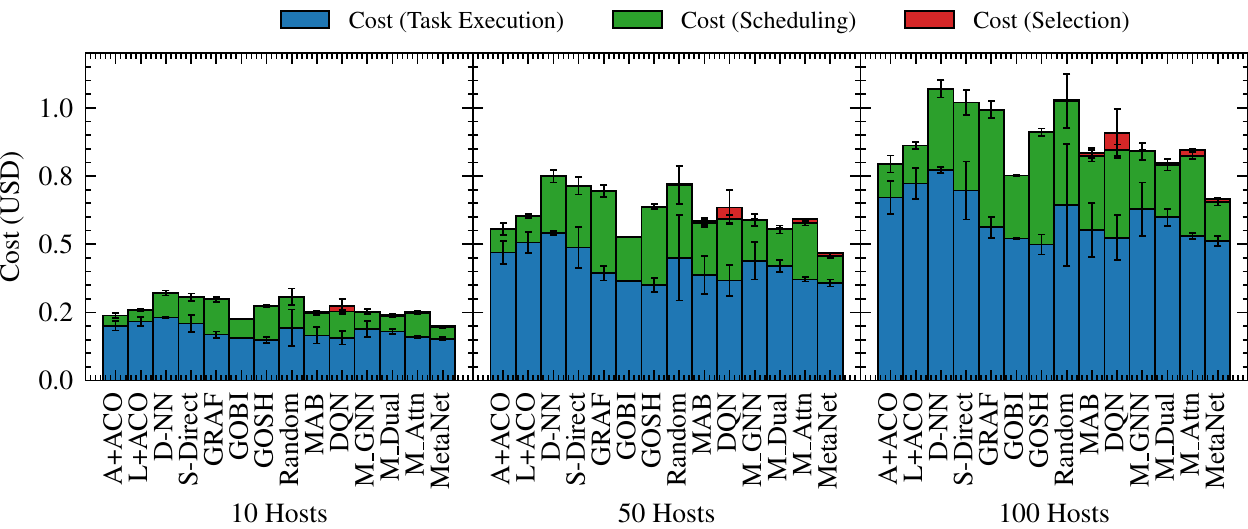}
    \caption{Execution cost (in USD) per task for different number of hosts in the cloud environment.}
    \label{fig:cost}
\end{figure}

\subsection{Baselines}
\label{sec:baselines}
We compare MetaNet against seven baselines, which also form our policy set $\mathcal{P}$. We integrate the ACO algorithm with two demand forecasting methods: AutoARIMA and LSTM, and call these ARIMA+ACO and LSTM+ACO. We also include other predict+optimize methods Decision-NN, Semi-Direct and GRAF. We also include state-of-the-art gradient-based optimization methods for scheduling, GOBI and GOSH (see Section~\ref{sec:related_work}). Thus, $\mathcal{P} = $ \{ARIMA+ACO, LSTM+ACO, Decision-NN, Semi-Direct, GRAF, GOBI, GOSH\}. This makes the size of the policy set $q = 7$. We also use a Multi-Armed Bandit (MAB) model using Upper-Confidence-Bound (UCB) exploration~\cite{kuleshov2014algorithms} and a Deep-Q-Network (DQN)~\cite{li2017deep} that dynamically choose a policy based on pre-trained models using data $\Lambda$ shown in~\eqref{eq:dataset}. The DQN baseline assume a Markov-Decision-Process (MDP) formulation wherein the state is defined as the policy that has been selected (a one-hot encoding corresponding to the selection) and action corresponds to choosing the same or a different policy at each scheduling interval. The MAB baseline, on the other hand, assumes a stateless input and just makes policy selection using the UCB approach. We also include a lightweight random policy selection method as a baseline indicating the importance of careful policy selection at the meta level.

\subsection{Comparison Metrics} 

We use the following evaluation metrics to test the efficacy of the MetaNet model as motivated from prior works~\cite{basu2019learn, tuli2020dynamic, tuli2021hunter, tuli2019fogbus, beloglazov2012optimal, mirhosseini2021parslo}.
\begin{enumerate}[leftmargin=*]
    \item \textit{Average Cost per task} which is given as 
    \[\frac{1}{T} \sum_{t=1}^T \phi^\pi_t\]
    such that $\pi$ is the policy used for generating the scheduling decision.
    
    \item \textit{Energy Consumption} of an experiment is given as the average energy consumed by all host machines in the cloud setup as \[\sum_{h^j \in H} \int_{t=1}^{T} Power(t, h^j) \cdot \Delta dt\] 
    where $Power_{h^j}(t)$ is the average power of host $h^j$ in interval $I_t$.
    
    \item \textit{Average Response Time} is the average response time of all workloads
    \[\frac{\sum_{t=1}^T \sum_{w^i_t \in \mathcal{W}_t} Response\ Time(w^i_t)}{\sum_{t=1}^T n_t}.\]
    
    \item \textit{Average Waiting Time} is the average time a task waits to get allocated to a host
    \[\frac{\sum_{t=1}^T \sum_{w^i_t \in \mathcal{W}_t} Waiting\ Time(w^i_t)}{\sum_{t=1}^T n_t},\]
    where waiting time is the sum of scheduler selection time (in case of MetaNet), scheduling time and container allocation time to the corresponding host. 
    
    \item \textit{SLA Violations} which is given as  
    \[\frac{\sum_{t=1}^T \sum_{w^i_t \in \mathcal{W}_t} \mathds{1}(Response\ Time(w^i_t) \leq \psi(w^i_t))}{\sum_{t=1}^T n_t},\] 
    where $\psi(w^i_t)$ is the 90$^{th}$ percentile response time for this application type (ResNet18, ResNet34, ResNext32x4d, SqueezeNet, GoogleNet, MobileNetV2, MnasNet) on the state of the art baseline GOSH. This percentile-based definition of the SLA deadline, inspired from~\cite{boloor2010dynamic, tuli2021cosco},  is defined for the response time metrics of completed tasks.
    
    \item \textit{Fairness} which is given as the Jain's Fairness Index 
    \[\frac{( \sum_{t=1}^T \sum_{w^i_t\in \mathcal{W}_t} Response\ Time(w^i_t))^2}{(\sum_{t=1}^T |\mathcal{W}_t|) \times ( \sum_{t=1}^T \sum_{w^i_t\in \mathcal{W}_t} Response\ Time(w^i_t)^2 )}.\]
    
    \item \textit{Average CPU Utilization} which is the average utilization of CPU in percentage of all hosts
    \[\frac{\sum_{t=1}^T \sum_{h^j \in \mathcal{H}} \bar{c}^j_t}{T \cdot m}.\]
\end{enumerate}

\subsection{Visualization of Dynamic Policy Selection}

Figure~\ref{fig:vis} visualizes the MetaNet approach running on the setup and workloads described above. The x-axis denotes the scheduling interval and the y-axis denotes the cost estimate $\phi^k_{max} \cdot \hat{\phi}^k_t + \omega^k_{max} \cdot \hat{\omega}^k_t$ for the three most frequently used policies in $\mathcal{P}$ for readability: ARIMA+ACO, GOBI and GOSH. The highlighted bands indicate the selected scheduling policy. The selected policy corresponds to the one with the least estimated cost. The cost estimates are non-stationary, further corroborating the need for dynamic selection of the scheduling policies in volatile workload and host setups. 

\subsection{Results}
Figure~\ref{fig:results} presents the QoS metrics for all baseline models and MetaNet for the Azure testbed with 100 hosts. Figure~\ref{fig:cost} shows the overall cost scores for a different number of hosts in the cloud environment. Across all metrics, MetaNet outperforms the baselines. The dynamic optimization method, DQN, has a high inference and fine-tuning time as it requires DNN inference for each scheduler instead of a single inference in MetaNet. This gives rise to higher selection costs for DQN compared to MetaNet and MAB approaches (see Figure~\ref{fig:cost}). GOSH achieves the lowest energy consumption across all baselines. MetaNet improves the energy consumption by allocating tasks, \textit{i.e.}, to the same hosts to minimize execution costs and consequently the active hosts in the system. This is shown in Fig.~\ref{fig:cpu} as the average CPU utilization of the MetaNet approach is the highest of 87.4\%. This enables MetaNet to give an average energy consumption of 1.561 KW-hr, 42.61\% lower than the lowest baseline value of 2.720 KW-hr by GOSH. This also allows MetaNet to provide an average cost of up to 11\% lower than the most cost-efficient approach, \textit{i.e.}, GOBI (see Fig.~\ref{fig:cost}. MetaNet also gives an average response time of 11.272 seconds, 8\% lower than the best average response time of 12.255 of the Decision-NN baseline. This is due to the lower wait times, as is observed from Figs.~\ref{fig:wait}. MetaNet also gives the lowest average SLA violation rate of 0.102, \textit{i.e.}, 13\% lower compared to the lowest baseline score of GOSH, which is 0.118. This is due to the lower average response time (see Fig.~\ref{fig:rt}). In terms of the fairness score, MetaNet gives the highest value of 0.893. Fig.~\ref{eq:cost} also demonstrates that as the number of host machines in the cloud environment increases, so does the cost gains from the MetaNet model. Dynamic optimization baselines perform poorly due to the stateless assumption in MAB that does not account for environment dynamism, and DQN being slow to adapt in volatile settings~\cite{tuli2021cosco}.

\subsection{Ablation Analysis}
To test the importance of the graph neural network, dual-headed prediction of cost and scheduling time and self-attention module, we modify the MetaNet neural network as follows. First, we consider a model without the output of the graph attention network, \textit{i.e.}, utilize a zero vector in the case of $E^S$. We call this the MetaNet\_GNN model. Second, instead of predicting the cost and scheduling time separately, we produce a single cost metric as the output of the neural network. We call this MetaNet\_Dual model. Third, we replace the self-attention module with a feed-forward network to test the importance of temporal trends that the transformer captures. We call this the MetaNet\_Attn model. The results in Figures~\ref{eq:cost} and~\ref{fig:results} show a drop in all performance metrics for these models when compared to MetaNet, demonstrating the effectiveness of the scheduling decision based graph inference, independent prediction of complementary metrics of execution cost with scheduling time, and the attention mechanism in the neural model.

\begin{figure}
    \centering
    \includegraphics[width=\linewidth]{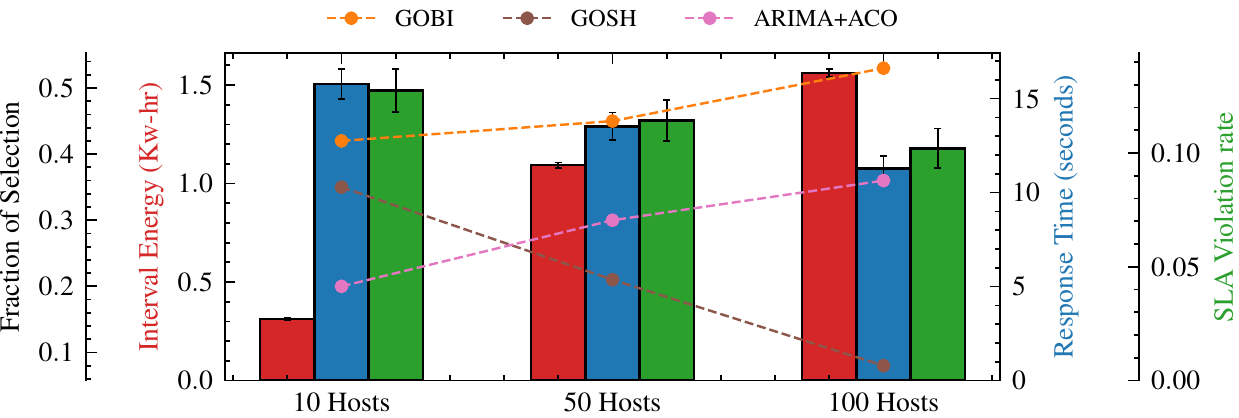}
    \caption{Sensitivity Analysis with number of cloud hosts.}
    \label{fig:sens}
\end{figure}

\subsection{Sensitivity Analysis}

Figure~\ref{fig:sens} shows the performance of the MetaNet model as we vary the number of worker nodes (hosts) in the cloud environment. it also shows the selection frequencies of two light-weight schedulers: GOBI and ARIMA+ACO and a relatively computationally heavier scheduler GOSH. Clearly, with an increasing number of hosts, energy consumption also increases. However, the response time and SLA violation rates decreases as the load over the host machines drops as more worker nodes are available. However, the figure shows that as the number of hosts increases, the fraction of times the GOSH method is selected drops significantly. This is due to the poor scalability of the GOSH method compared to ARIMA+ACO. This is because the GOSH approach performs higher-order optimization over the scheduling decision. In such cases, the MetaNet approach finds the GOSH scheduler not worth the additional costs due to higher scheduling time. It thus selects more straightforward methods such as GOBI and ARIMA+ACO more frequently in such a case.

\section{Conclusions}
\label{sec:conclusion}

This paper proposes MetaNet, a surrogate model-based solution to dynamically select the cost-optimal scheduler at each scheduling interval in a cloud computing environment. To do this, it predicts estimates of the task execution costs as well as the scheduling time for a set of scheduling policies. Such a surrogate allows us to compute an estimate of the total cost implications of running a scheduling policy without executing all policies and select the one with the least cost estimate. Periodic fine-tuning of the neural network surrogate model facilitates adaptation in dynamic workload settings. Experiments with real-life AI-based benchmark applications on a public cloud testbed show that MetaNet gives upto overall 11\% lower operational costs, 43\% lower energy consumption, 8\% lower average response time and 13\% lower SLA violation rates compared to state-of-the-art methods. Being policy agnostic, this method can be updated with the latest scheduling policies or even extended to other types of resource management problems or computing paradigms in the future~\cite{}.

\section*{Software Availability}
The code is publicly available on GitHub under BSD-3 licence at \url{https://github.com/imperial-qore/MetaNet}.

\section*{Acknowledgments}
Shreshth Tuli is supported by the President’s Ph.D. Scholarship at the Imperial College London. We thank Shikhar Tuli for constructive discussions. 

\bibliographystyle{IEEEtran}
\bibliography{references}

\begin{thebibliography}{10}
\providecommand{\url}[1]{#1}
\csname url@samestyle\endcsname
\providecommand{\newblock}{\relax}
\providecommand{\bibinfo}[2]{#2}
\providecommand{\BIBentrySTDinterwordspacing}{\spaceskip=0pt\relax}
\providecommand{\BIBentryALTinterwordstretchfactor}{4}
\providecommand{\BIBentryALTinterwordspacing}{\spaceskip=\fontdimen2\font plus
\BIBentryALTinterwordstretchfactor\fontdimen3\font minus
  \fontdimen4\font\relax}
\providecommand{\BIBforeignlanguage}[2]{{%
\expandafter\ifx\csname l@#1\endcsname\relax
\typeout{** WARNING: IEEEtran.bst: No hyphenation pattern has been}%
\typeout{** loaded for the language `#1'. Using the pattern for}%
\typeout{** the default language instead.}%
\else
\language=\csname l@#1\endcsname
\fi
#2}}
\providecommand{\BIBdecl}{\relax}
\BIBdecl

\bibitem{gill2019transformative}
S.~S. Gill, S.~Tuli, M.~Xu, I.~Singh, K.~V. Singh \emph{et~al.},
  ``{Transformative effects of IoT, Blockchain and Artificial Intelligence on
  cloud computing: Evolution, vision, trends and open challenges},''
  \emph{Internet of Things}, vol.~8, pp. 100--118, 2019.

\bibitem{boudi2021ai}
A.~Boudi, M.~Bagaa, P.~P{\"o}yh{\"o}nen, T.~Taleb, and H.~Flinck, ``{AI-based
  resource management in beyond 5G cloud native environment},'' \emph{IEEE
  Network}, vol.~35, no.~2, pp. 128--135, 2021.

\bibitem{tuli2021mcds}
S.~Tuli, G.~Casale, and N.~R. Jennings, ``{MCDS: AI Augmented Workflow
  Scheduling in Mobile Edge Cloud Computing Systems},'' \emph{IEEE Transactions
  on Parallel and Distributed Systems}, 2022.

\bibitem{cai2016iot}
H.~Cai, B.~Xu, L.~Jiang, and A.~V. Vasilakos, ``Iot-based big data storage
  systems in cloud computing: perspectives and challenges,'' \emph{IEEE
  Internet of Things Journal}, vol.~4, no.~1, pp. 75--87, 2016.

\bibitem{decisionnn}
B.~Wilder, B.~Dilkina, and M.~Tambe, ``Melding the data-decisions pipeline:
  Decision-focused learning for combinatorial optimization,'' in \emph{AAAI},
  vol.~33, no.~01, 2019, pp. 1658--1665.

\bibitem{tuli2021hunter}
S.~Tuli, S.~S. Gill, M.~Xu, P.~Garraghan, R.~Bahsoon \emph{et~al.}, ``{HUNTER:
  AI based holistic resource management for sustainable cloud computing},''
  \emph{Journal of Systems and Software}, pp. 111--124, 2021.

\bibitem{semidirect}
P.~J. Stuckey, T.~Guns, J.~Bailey, C.~Leckie, K.~Ramamohanarao \emph{et~al.},
  ``Dynamic programming for predict+optimise,'' in \emph{AAAI}, vol.~34,
  no.~02, 2020, pp. 1444--1451.

\bibitem{tuli2021cosco}
S.~Tuli, S.~R. Poojara, S.~N. Srirama, G.~Casale, and N.~R. Jennings, ``{COSCO:
  Container Orchestration Using Co-Simulation and Gradient Based Optimization
  for Fog Computing Environments},'' \emph{IEEE Transactions on Parallel and
  Distributed Systems}, vol.~33, no.~1, pp. 101--116, 2022.

\bibitem{graf}
J.~Park, B.~Choi, C.~Lee, and D.~Han, ``Graf: a graph neural network based
  proactive resource allocation framework for slo-oriented microservices,'' in
  \emph{Proceedings of the 17th International Conference on emerging Networking
  EXperiments and Technologies}, 2021, pp. 154--167.

\bibitem{tuli2021gosh}
S.~Tuli, G.~Casale, and N.~R. Jennings, ``{GOSH: Task Scheduling using Deep
  Surrogate Models in Fog Computing Environments},'' \emph{IEEE Transactions on
  Parallel and Distributed Systems}, 2022.

\bibitem{tuli2019fogbus}
S.~Tuli, R.~Mahmud, S.~Tuli, and R.~Buyya, ``Fogbus: A blockchain-based
  lightweight framework for edge and fog computing,'' \emph{Journal of Systems
  and Software}, 2019.

\bibitem{alkhanak2015cost}
E.~N. Alkhanak, S.~P. Lee, and S.~U.~R. Khan, ``Cost-aware challenges for
  workflow scheduling approaches in cloud computing environments: Taxonomy and
  opportunities,'' \emph{Future Generation Computer Systems}, vol.~50, pp.
  3--21, 2015.

\bibitem{tuli2021pregan}
S.~Tuli, G.~Casale, and N.~R. Jennings, ``{PreGAN: Preemptive Migration
  Prediction Network for Proactive Fault-Tolerant Edge Computing},'' in
  \emph{IEEE Conference on Computer Communications (INFOCOM)}.\hskip 1em plus
  0.5em minus 0.4em\relax IEEE, 2022.

\bibitem{traverso2013temporal}
S.~Traverso, M.~Ahmed, M.~Garetto, P.~Giaccone, E.~Leonardi \emph{et~al.},
  ``Temporal locality in today's content caching: Why it matters and how to
  model it,'' \emph{ACM SIGCOMM Computer Communication Review}, vol.~43, no.~5,
  pp. 5--12, 2013.

\bibitem{tuli2022tranad}
S.~Tuli, G.~Casale, and N.~R. Jennings, ``{TranAD: Deep Transformer Networks
  for Anomaly Detection in Multivariate Time Series Data},'' \emph{arXiv
  preprint arXiv:2201.07284}, 2022.

\bibitem{chen2021deep}
Y.~Chen and G.~Casale, ``Deep learning models for automated identification of
  scheduling policies,'' in \emph{2021 29th International Symposium on
  Modeling, Analysis, and Simulation of Computer and Telecommunication Systems
  (MASCOTS)}.\hskip 1em plus 0.5em minus 0.4em\relax IEEE, 2021, pp. 1--8.

\bibitem{tuli2022metanet}
S.~Tuli, G.~Casale, and N.~R. Jennings, ``{Learning to Dynamically Select the
  Optimal Scheduler in Cloud Computing Environments},'' \emph{SIGMETRICS
  Perform. Eval. Rev.}, 2022.

\bibitem{uahs}
C.~Luo, B.~Qiao, X.~Chen, P.~Zhao, R.~Yao \emph{et~al.}, ``Intelligent virtual
  machine provisioning in cloud computing,'' in \emph{IJCAI}, 2020, pp.
  1495--1502.

\bibitem{hyndman2018forecasting}
R.~J. Hyndman and G.~Athanasopoulos, \emph{Forecasting: principles and
  practice}.\hskip 1em plus 0.5em minus 0.4em\relax OTexts, 2018.

\bibitem{chen2018resource}
J.~Chen and Y.~Wang, ``A resource demand prediction method based on eemd in
  cloud computing,'' \emph{Procedia Computer Science}, vol. 131, pp. 116--123,
  2018.

\bibitem{arima}
P.~Singh, P.~Gupta, and K.~Jyoti, ``{TASM: Technocrat ARIMA and SVR model for
  workload prediction of web applications in cloud},'' \emph{Cluster
  Computing}, vol.~22, no.~2, pp. 619--633, 2019.

\bibitem{lstm}
S.~Ouhame, Y.~Hadi, and A.~Ullah, ``An efficient forecasting approach for
  resource utilization in cloud data center using cnn-lstm model,''
  \emph{Neural Computing and Applications}, pp. 1--13, 2021.

\bibitem{aco}
M.~Aliyu, M.~Murali, A.~Y. Gital, and S.~Boukari, ``Efficient metaheuristic
  population-based and deterministic algorithm for resource provisioning using
  ant colony optimization and spanning tree,'' \emph{International Journal of
  Cloud Applications and Computing (IJCAC)}, vol.~10, no.~2, pp. 1--21, 2020.

\bibitem{cahs}
C.~Luo, B.~Qiao, W.~Xing, X.~Chen, P.~Zhao \emph{et~al.}, ``Correlation-aware
  heuristic search for intelligent virtual machine provisioning in cloud
  systems,'' in \emph{AAAI}, vol.~35, no.~14, 2021, pp. 12\,363--12\,372.

\bibitem{nandi2001artificial}
S.~Nandi, S.~Ghosh, S.~S. Tambe, and B.~D. Kulkarni, ``Artificial
  neural-network-assisted stochastic process optimization strategies,''
  \emph{AIChE journal}, vol.~47, no.~1, pp. 126--141, 2001.

\bibitem{tuli2022carol}
S.~Tuli, G.~Casale, and N.~R. Jennings, ``{CAROL: Confidence-Aware Resilience
  Model for Edge Federations},'' in \emph{IEEE/IFIP International Conference on
  Dependable Systems and Networks (DSN)}.\hskip 1em plus 0.5em minus
  0.4em\relax IEEE, 2022.

\bibitem{nninversion}
E.~Wong and J.~Z. Kolter, ``Neural network inversion beyond gradient descent,''
  \emph{Advances in Neural Information Processing Systems, Workshop on
  Optimization for Machine Learning}, 2017.

\bibitem{basu2019learn}
D.~Basu, X.~Wang, Y.~Hong, H.~Chen, and S.~Bressan, ``Learn-as-you-go with
  megh: Efficient live migration of virtual machines,'' \emph{IEEE Transactions
  on Parallel and Distributed Systems}, vol.~30, no.~8, pp. 1786--1801, 2019.

\bibitem{gat}
P.~Veli{\v{c}}kovi{\'{c}}, G.~Cucurull, A.~Casanova, A.~Romero, P.~Li{\`{o}}
  \emph{et~al.}, ``{Graph Attention Networks},'' \emph{International Conference
  on Learning Representations}, 2018.

\bibitem{xie2020mgat}
Y.~Xie, Y.~Zhang, M.~Gong, Z.~Tang, and C.~Han, ``Mgat: Multi-view graph
  attention networks,'' \emph{Neural Networks}, vol. 132, pp. 180--189, 2020.

\bibitem{vaswani2017attention}
A.~Vaswani, N.~Shazeer, N.~Parmar, J.~Uszkoreit, L.~Jones \emph{et~al.},
  ``Attention is all you need,'' in \emph{Proceedings of the 31st International
  Conference on Neural Information Processing Systems}, 2017, pp. 6000--6010.

\bibitem{karlik2011performance}
B.~Karlik and A.~V. Olgac, ``Performance analysis of various activation
  functions in generalized mlp architectures of neural networks,''
  \emph{International Journal of Artificial Intelligence and Expert Systems},
  vol.~1, no.~4, pp. 111--122, 2011.

\bibitem{alghushairy2020review}
O.~Alghushairy, R.~Alsini, T.~Soule, and X.~Ma, ``A review of local outlier
  factor algorithms for outlier detection in big data streams,'' \emph{Big Data
  and Cognitive Computing}, vol.~5, no.~1, p.~1, 2020.

\bibitem{saleh2019dynamic}
N.~Saleh and M.~Mashaly, ``A dynamic simulation environment for container-based
  cloud data centers using containercloudsim,'' in \emph{2019 Ninth
  International Conference on Intelligent Computing and Information Systems
  (ICICIS)}.\hskip 1em plus 0.5em minus 0.4em\relax IEEE, 2019, pp. 332--336.

\bibitem{han2015learning}
S.~Han, J.~Pool, J.~Tran, and W.~Dally, ``Learning both weights and connections
  for efficient neural network,'' \emph{Advances in neural information
  processing systems}, vol.~28, 2015.

\bibitem{pan2009survey}
S.~J. Pan and Q.~Yang, ``A survey on transfer learning,'' \emph{IEEE
  Transactions on knowledge and data engineering}, vol.~22, no.~10, pp.
  1345--1359, 2009.

\bibitem{grinberg2018flask}
M.~Grinberg, \emph{Flask web development: developing web applications with
  python}.\hskip 1em plus 0.5em minus 0.4em\relax " O'Reilly Media, Inc.",
  2018.

\bibitem{venkatesh2019fast}
R.~S. Venkatesh, T.~Smejkal, D.~S. Milojicic, and A.~Gavrilovska, ``Fast
  in-memory criu for docker containers,'' in \emph{The International Symposium
  on Memory Systems}, 2019, pp. 53--65.

\bibitem{casale2020radon}
G.~Casale, M.~Arta{\v{c}}, W.-J. van~den Heuvel, A.~van Hoorn, P.~Jakovits
  \emph{et~al.}, ``Radon: rational decomposition and orchestration for
  serverless computing,'' \emph{SICS Software-Intensive Cyber-Physical
  Systems}, vol.~35, no.~1, pp. 77--87, 2020.

\bibitem{tuli2019edgelens}
S.~Tuli, N.~Basumatary, and R.~Buyya, ``Edgelens: Deep learning based object
  detection in integrated iot, fog and cloud computing environments,''
  \emph{arXiv preprint arXiv:1906.11056}, 2019.

\bibitem{luo2018aiot}
C.~Luo, F.~Zhang, C.~Huang, X.~Xiong, J.~Chen \emph{et~al.}, ``Aiot bench:
  towards comprehensive benchmarking mobile and embedded device intelligence,''
  in \emph{International Symposium on Benchmarking, Measuring and
  Optimization}.\hskip 1em plus 0.5em minus 0.4em\relax Springer, 2018, pp.
  31--35.

\bibitem{coco}
T.-Y. Lin, M.~Maire, S.~Belongie, J.~Hays, P.~Perona \emph{et~al.}, ``Microsoft
  coco: Common objects in context,'' in \emph{European conference on computer
  vision}.\hskip 1em plus 0.5em minus 0.4em\relax Springer, 2014, pp. 740--755.

\bibitem{mao2016dynamic}
Y.~Mao, J.~Zhang, and K.~B. Letaief, ``Dynamic computation offloading for
  mobile-edge computing with energy harvesting devices,'' \emph{IEEE Journal on
  Selected Areas in Communications}, vol.~34, no.~12, pp. 3590--3605, 2016.

\bibitem{gilly2020modelling}
K.~Gilly, S.~Alcaraz, N.~Aknin, S.~Filiposka, and A.~Mishev, ``Modelling edge
  computing in urban mobility simulation scenarios,'' in \emph{2020 IFIP
  Networking Conference (Networking)}.\hskip 1em plus 0.5em minus 0.4em\relax
  IEEE, 2020, pp. 539--543.

\bibitem{krajzewicz2012recent}
D.~Krajzewicz, J.~Erdmann, M.~Behrisch, and L.~Bieker, ``Recent development and
  applications of sumo-simulation of urban mobility,'' \emph{International
  journal on advances in systems and measurements}, vol.~5, no. 3\&4, 2012.

\bibitem{kuleshov2014algorithms}
V.~Kuleshov and D.~Precup, ``Algorithms for multi-armed bandit problems,''
  \emph{arXiv:1402.6028}, 2014.

\bibitem{li2017deep}
Y.~Li, ``Deep reinforcement learning: An overview,'' \emph{arXiv:1701.07274},
  2017.

\bibitem{tuli2020dynamic}
S.~Tuli, S.~Ilager, K.~Ramamohanarao, and R.~Buyya, ``{Dynamic Scheduling for
  Stochastic Edge-Cloud Computing Environments using A3C learning and Residual
  Recurrent Neural Networks},'' \emph{IEEE Transactions on Mobile Computing},
  2020.

\bibitem{beloglazov2012optimal}
A.~Beloglazov and R.~Buyya, ``Optimal online deterministic algorithms and
  adaptive heuristics for energy and performance efficient dynamic
  consolidation of virtual machines in cloud data centers,'' \emph{Concurrency
  and Computation: Practice and Experience}, vol.~24, no.~13, pp. 1397--1420,
  2012.

\bibitem{mirhosseini2021parslo}
A.~Mirhosseini, S.~Elnikety, and T.~F. Wenisch, ``Parslo: A gradient
  descent-based approach for near-optimal partial slo allotment in
  microservices,'' in \emph{Proceedings of the ACM Symposium on Cloud
  Computing}, 2021, pp. 442--457.

\bibitem{boloor2010dynamic}
K.~Boloor, R.~Chirkova, Y.~Viniotis, and T.~Salo, ``Dynamic request allocation
  and scheduling for context aware applications subject to a percentile
  response time sla in a distributed cloud,'' in \emph{2010 IEEE Second
  International Conference on Cloud Computing Technology and Science}.\hskip
  1em plus 0.5em minus 0.4em\relax IEEE, 2010, pp. 464--472.

\end{thebibliography}

\end{document}